\documentclass[aps,onecolumn]{revtex4}
\usepackage{morefloats}
\usepackage{float}
\usepackage{graphicx}
\usepackage{dcolumn}
\usepackage{amsmath}
\usepackage{subfigure}
\usepackage{epsfig}
\input{epsf}

\begin{document}


\title{Coupling of PbS Quantum Dots to Photonic Crystal Cavities at Room Temperature}


\author{Ilya Fushman}
\altaffiliation[Also at: ]{Department of Biophysics, Stanford
University, Stanford, CA 94305}
\email[]{ifushman@stanford.edu}
\author{Dirk Englund}
\altaffiliation[Also at: ]{Department of Applied Physics, Stanford
University, Stanford, CA 94305}
\author{Jelena Vu\v{c}kovi\'{c}}
\altaffiliation[Also at: ]{Department of Electrical Engineering,
Stanford University, Stanford, CA 94305}




\date{\today}

\begin{abstract}

We demonstrate the coupling of PbS quantum dot emission to photonic crystal cavities at room temperature. The cavities are defined in 33\% Al, AlGaAs membranes on top of oxidized AlAs. Quantum dots were dissolved in Poly-methyl-methacrylate (PMMA) and spun on top of the cavities. Quantum dot emission is shown to map out the structure resonances, and may prove to be viable sources for room temperature cavity coupled single photon generation for quantum information processing applications. These results also indicate that such commercially available quantum dots can be used for passive structure characterization. The deposition technique is versatile and allows layers with different dot densities and emission wavelengths to be re-deposited on the same chip.

\end{abstract}

\maketitle
\clearpage
Photonic crystal cavities are a very suitable medium for control and confinement of light on scales on the order of a cubic wavelength of light in the material. The small modal volumes and high quality factors for these cavities make them strong candidates for efficient single photon generation for quantum information processing \cite{ref:Vuckovic2003,ref:Dirk2005}, low threshold lasing \cite{ref:Yablonovich1993_PCreview}, and nonlinear optics phenomena \cite{ref:Joannopoulos2004nonlinPC}. Our main motivation is the search for emitters that can be coupled to photonic crystal (PC) cavities in order to satisfy the constraints on indistinguishability and efficiency of the photon source for quantum information processing. Additional motivation is the characterization of the electromagnetic properties of passive elements to be used in photonic circuits. To date, successful generation of single photons coupled to PC microcavities has been demonstrated at cryogenic temperatures because the emitters of choice degrade with an increase in temperature. The emission wavelengths of the emitters have to be compatible with typical semiconductor materials used for microcavities \cite{ref:Dirk2005}. Room temperature generation of single photons has been observed from single molecules \cite{ref:Moerner2000SM}, nitrogen vacancy centers \cite{ref:Grangier2001NV}, and CdSe quantum dots \cite{ref:Michler2000sps_CdSe}, but collection efficiencies were reduced due to lack of coupling to good cavities. These sources, with the exception of nitrogen vacancy centers, operate at visible wavelengths and are therefore difficult to combine with cavities made in high index semiconductor slabs. In contrast, PbS quantum dots (and other colloidal quantum dots such as PbSe) can be made to cover a very broad wavelength range. Furthermore, these dots can be easily deposited onto passive structures in a low index polymer and can be used to map out the resonances of the structures. This method is much easier than transmission and reflectivity type measurements on single cavities. Lastly, PbS quantum dots are successfully used as fluorescent labels in biological imaging applications. Cavity enhanced emission and collection efficiency from such dots could prove to be a valuable technique for targeted signal amplification of a target molecule bound to or in the vicinity of the resonant cavity.

\begin{figure}
\centering
	\includegraphics[height=2in]{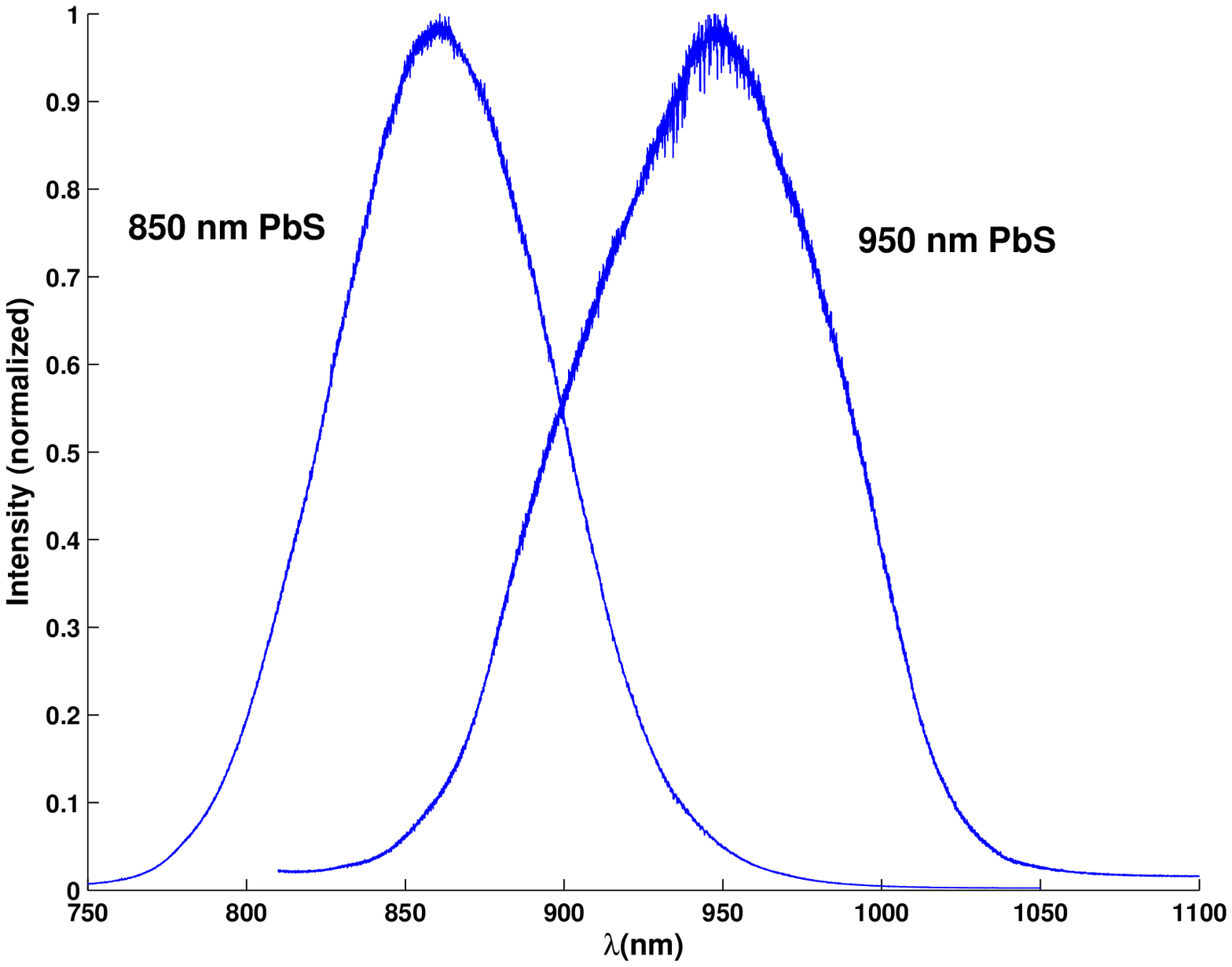}
	\caption{} \label{fig:PbS_Spectra}
\end{figure}

\begin{figure}
    \includegraphics[height=2in]{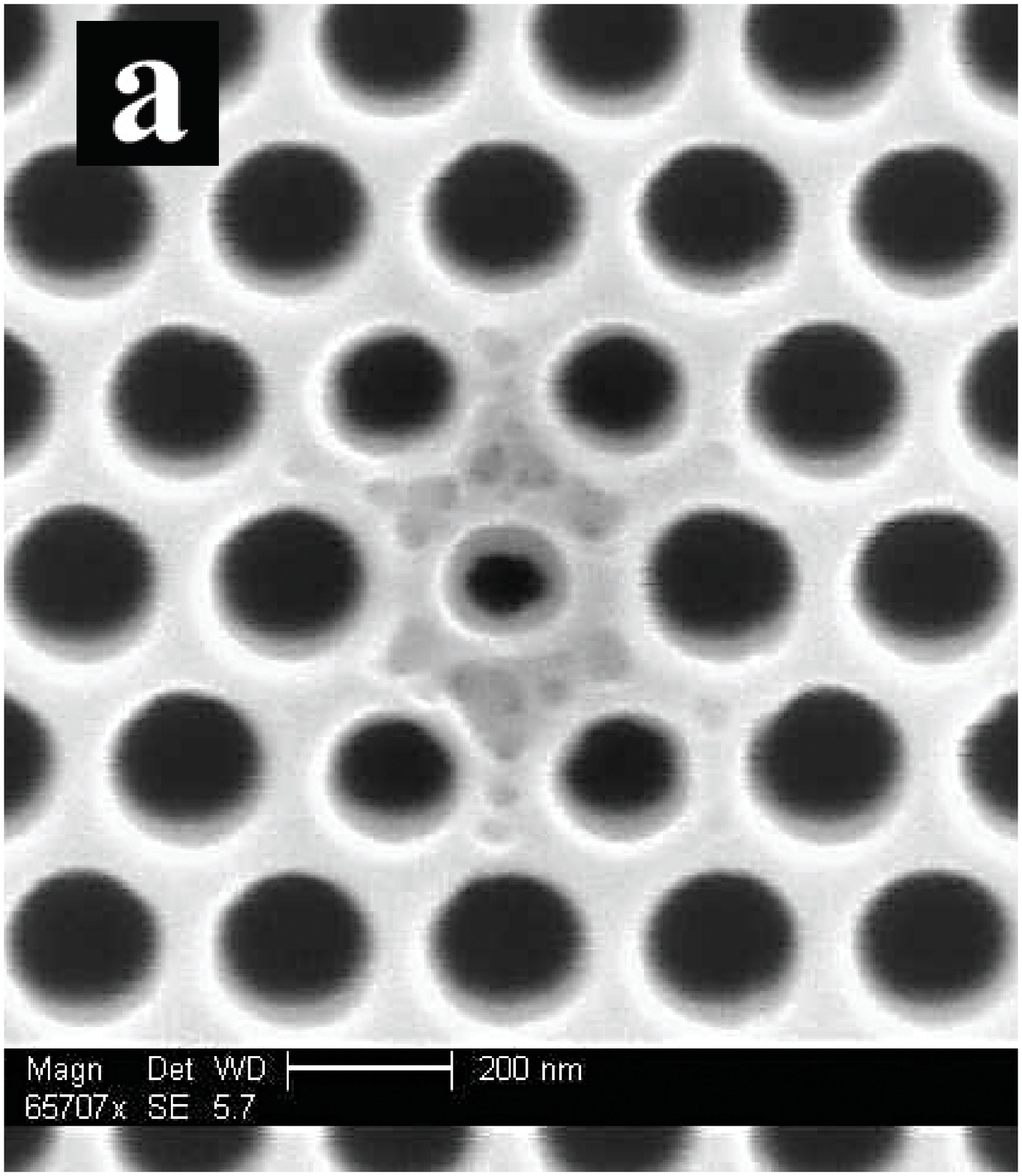}
		\includegraphics[height=2in]{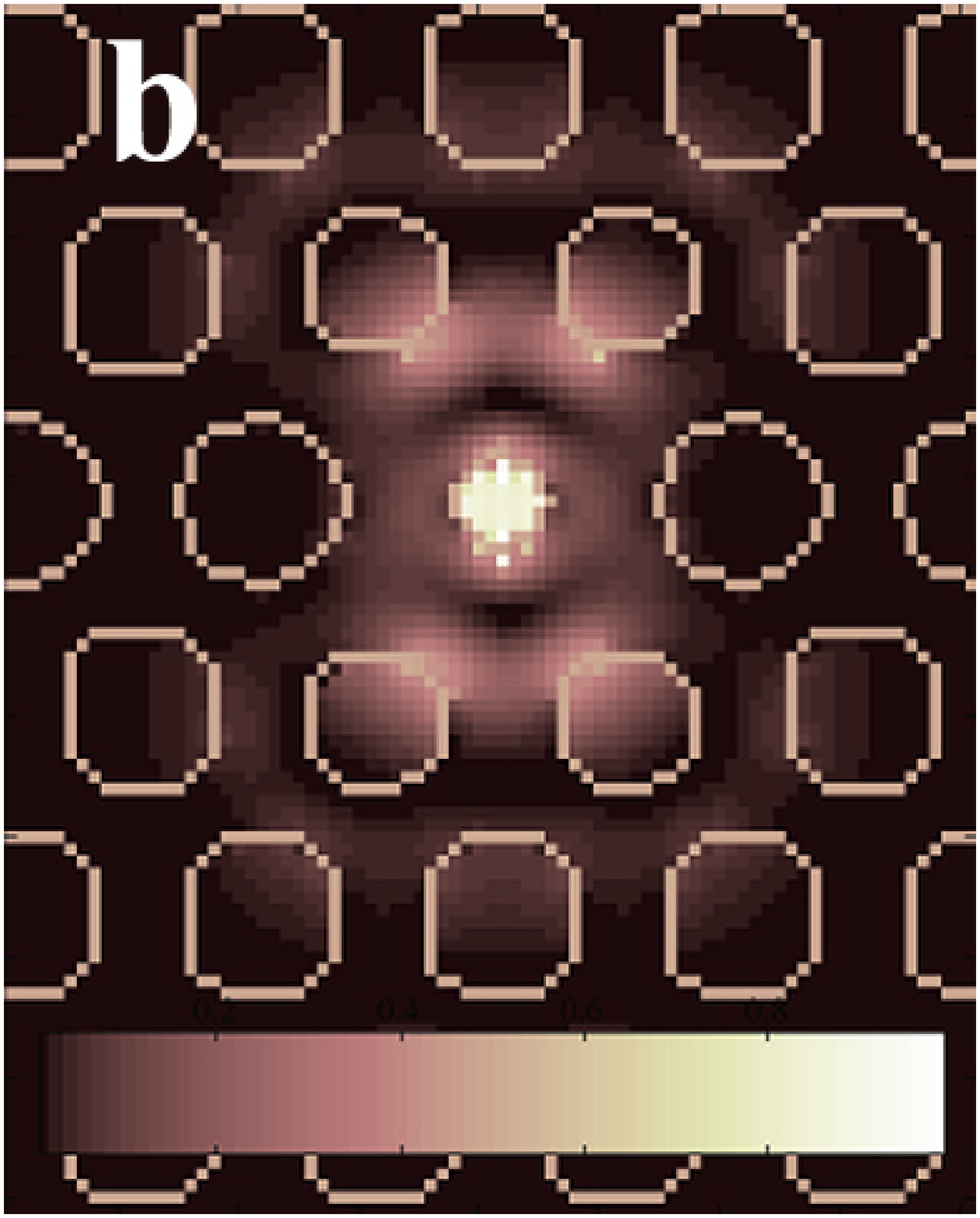}
		\includegraphics[height=2in]{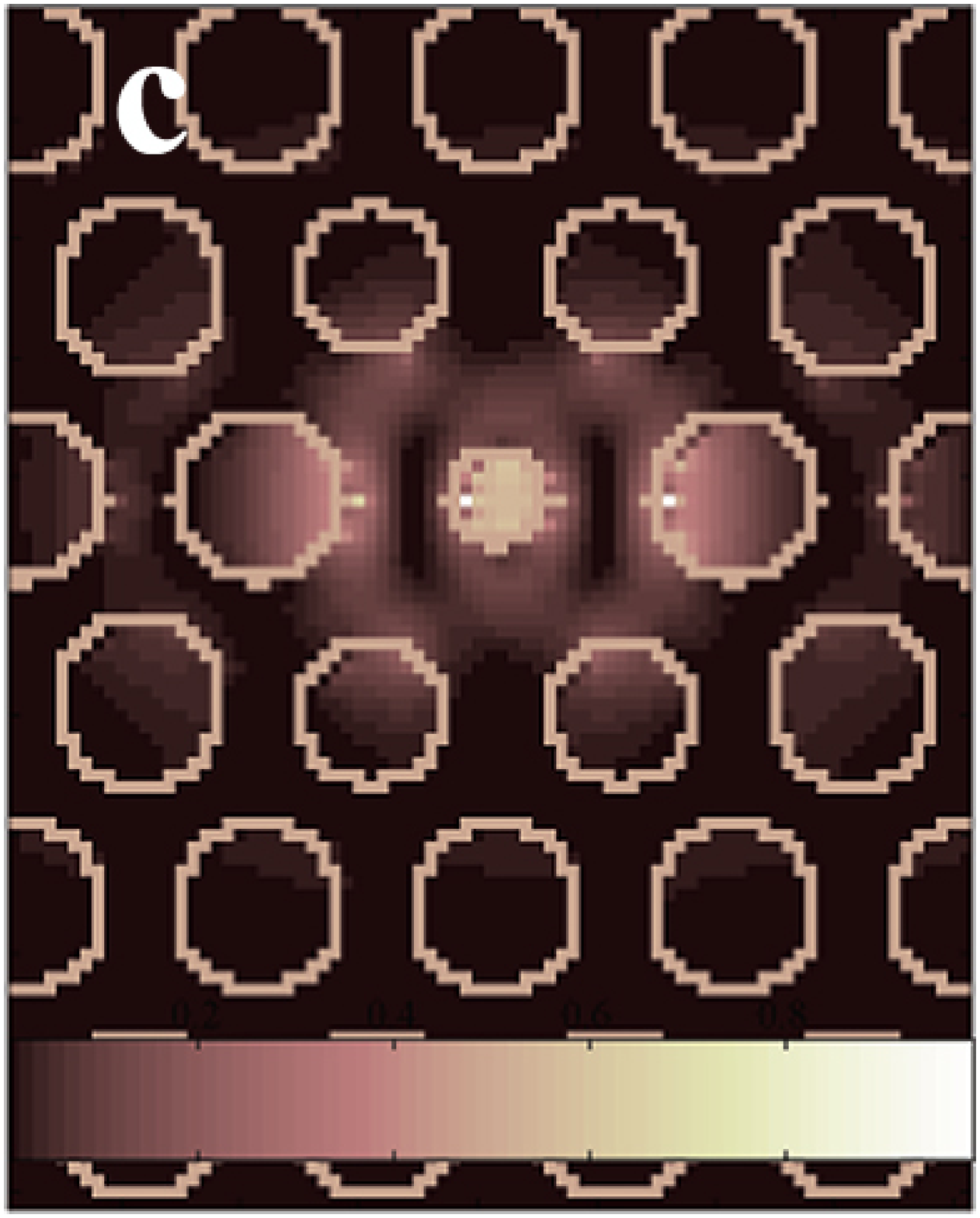}
\caption{} \label{fig:SEMCAV}
\end{figure}

\begin{figure}
    \includegraphics[height=2in]{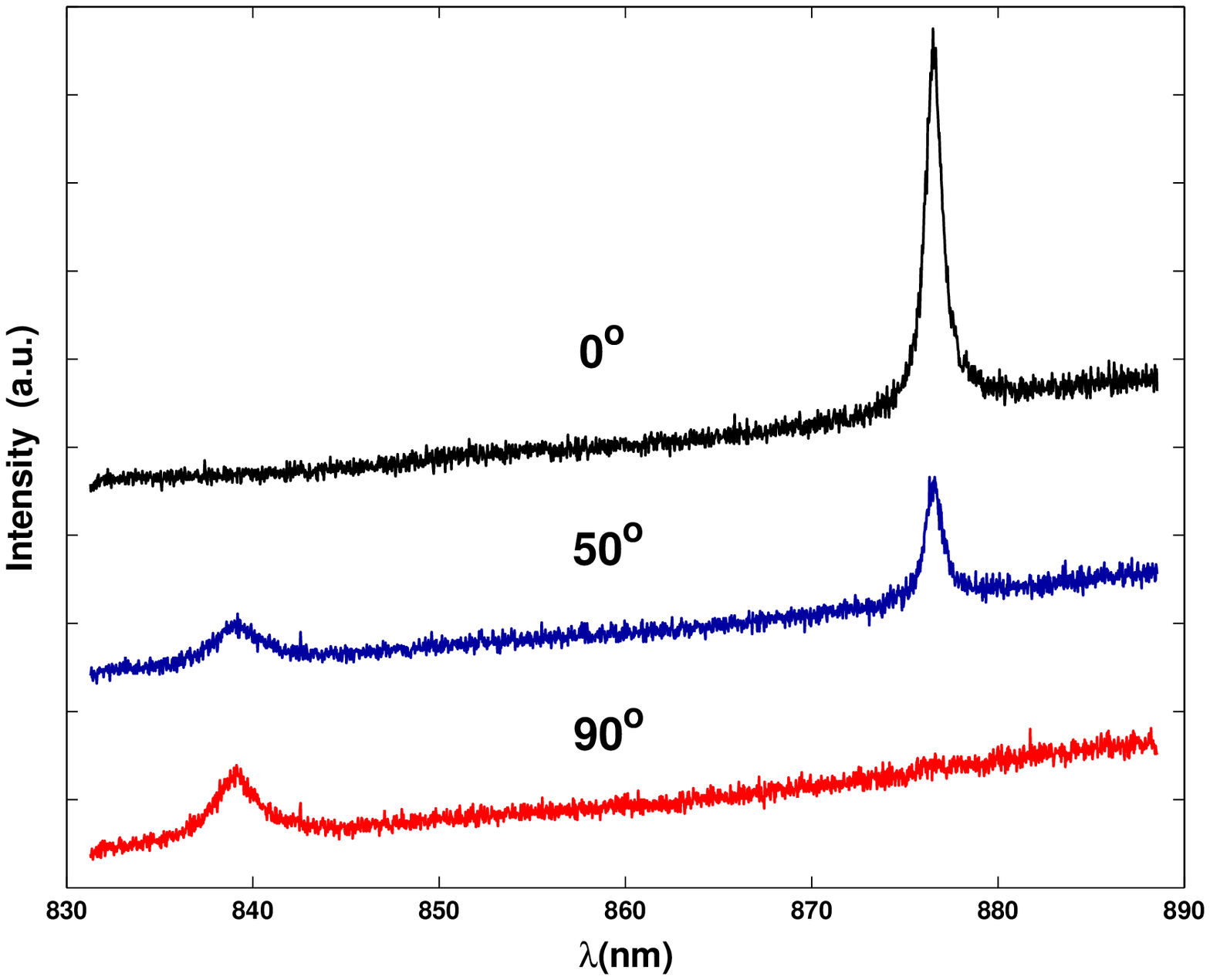}
		\includegraphics[height=2in]{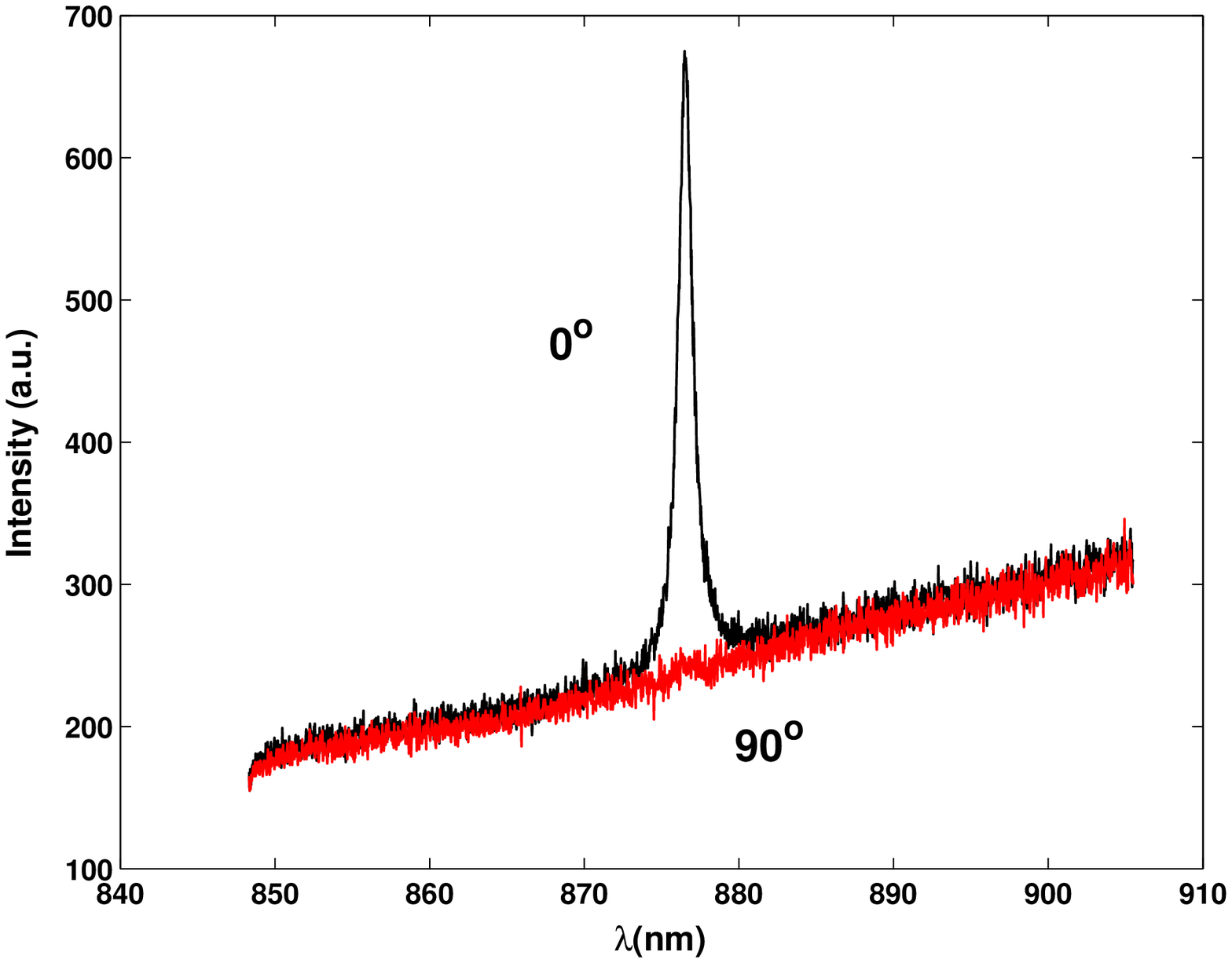}
\caption{} \label{fig:polspec}
\end{figure}

Here, we report on the coupling of emission from PbS quantum dots to semiconductor photonic crystal cavities at room temperature. Photonic crystal cavities were made in a 160 nm thick AlGaAs (33\% Al) membrane on top of 500 nm of AlAs. The cavities were defined in 3\% 450 K molecular weight (KmW) PMMA with the Raith Electron Beam Lithography system. The patterns were transferred from PMMA developed in 3:1 Methyl Isobutyl Ketone (MIBK) to the membrane with an electron-cyclotron resonance (ECR) plasma etch process. The Al rich substrate was then oxidized at $420^{o}$ C for 10 minutes in the presence of water vapor in order to create an index contrast of 3.4:1.8 between membrane and substrate. A scanning electron micrograph of the sample structure is shown in Fig. \ref{fig:SEMCAV}. PbS quantum dots emitting at 850 nm and 950 nm were obtained from Evident Technologies. The dots were dissolved together in 1\% 75 KmW PMMA at a concentration of 0.1 mg/ml in toluene and spun onto the structures at 2 krpm, resulting in a ~ 20nm-100nm membrane (bulk spectra shown in Fig. \ref{fig:PbS_Spectra}). The coating was not uniform due to the small chip size $\sim (5 mm^2)$ and presence of structures on the chip surface. This concentration of emitters corresponds to $\sim 10^{3}$  dots per $\mu{m}^2$. The dots were kept under vacuum, and excited with femtosecond pulses from a Ti:Saph laser at 760 nm. Cavity spectra, collected under pulsed excitation, are shown in Fig. \ref{fig:polspec}. The spectra reveal two orthogonally polarized modes as expected for this type of photonic crystal cavity. The linearly increasing background observed on the spectra can be explained by considering the emission profile of the dots and the reflectivity of the PMMA/AlGaAs/AlOx structure, which leads to an almost linear emission profile of dots on the AlGaAs slab.   

The cavity used is shown in Fig. \ref{fig:SEMCAV}. The coupling between a cavity and emitter depends on both the spatial alignment and the orientation of the emitter dipole moment relative to the cavity field. The chosen cavities support two dipole type (electric field primarily polarized in x and y) cavity modes with maxima in the central hole in (Fig. \ref{fig:SEMCAV}). The asymmetric cavity was modeled using Finite Difference Time Domain (FDTD) methods \cite{ref:Vuckovic2003}. The simulations predict the splitting between the two resonant peaks to be $\approx$ 43 nm, which is somewhat larger than the observed splitting of 37.5 nm and is likely due to deviations of fabricated structures from simulated ones. The quality factor (Q) is the figure of merit for these cavities, and can be written as  
\begin{equation}
\label{eq:Qtot} \frac{1}{Q} \equiv \frac{1}{Q_{||}} + \frac{1}{Q_\bot}
\end{equation}
The above description separates the cavity losses into an out-of-plane loss which leads to directly observed emission, and losses into the Bragg mirrors in the plane of the two dimensional crystal. The cavities investigated here have experimentally observed x-dipole Q's $\approx$ 400 and y-dipole Q's $\approx$ 200. Simulations for the asymmetric cavity predict that the out of plane $Q_\bot$ $\approx$ 8000 for the x-dipole mode and $\approx$ 4000 for the y-dipole. The in-plane $Q_{||}$ was only $\approx$ 450 and 400 for these modes, and thus limits the total Q. The center hole sidewalls are not vertical because of a reduced etch rate for smaller features and a lower dose delivered to the PMMA due to proximity effects. Since this hole is in the region of maximal electric field, imperfections will strongly reduce the quality factor. The two cavity modes are shown in Fig. \ref{fig:SEMCAV} b. In a perfectly symmetric cavity, where the two holes above and below the defect are unperturbed, the x and y modes are degenerate. We shift the four holes in order to increase the Q factor for the x dipole mode, as has been discussed in \cite{ref:Vuckovic2003}.   
	
The photonic crystal can both enhance and reduce spontaneous emission (SE) rates for the quantum dots. The enhancement is desirable for photon generation and spectroscopy applications due to increased signal rate and strength. The figure of merit for this modification is the Purcell factor $ F = \frac{\Gamma_{C}}{ \Gamma_{0}}$. Here $\Gamma$ is the emission rate in the cavity, and $\Gamma_0$ is that without the cavity. The observed spontaneous emission rate for an emitter with dipole moment $\vec{\mu}$ at a point on the cavity can be derived from Fermi's Golden rule \cite{ref:SCULLY}, and the total rate takes the form
\begin{equation}
\label{eq:Fcav} \frac{\Gamma_{C}}{ \Gamma_{0}} = F_{C}\left(\frac{\vec{E}\cdot\vec{\mu}}{\left|\max\{\vec{E}\}\right|\left|\vec{\mu}\right|}\right)^2 \frac{\Delta\lambda_c^2}{\Delta\lambda_c^2+4(\lambda-\lambda_c)^2}+F_{PC}
\end{equation}
Here $\vec{E}$ is the electric field at the position of the emitter, $V_{mode}$ is the cavity mode volume, $\Delta\lambda$ is the detuning from the cavity resonance wavelength $\lambda_c$. For an emitter that is on resonance with the cavity and has a dipole moment aligned with the field, $F_C = \left(\frac{3\lambda_{c}^3 Q}{4\pi^2 n^3 V_{mode}}\right)$. The mode volume is calculated from the FDTD simulation results as 
$V_{mode}=\int{\frac{  \epsilon\left|\vec{E}\right|^2}{\max\{\epsilon\left|\vec{E}\right|^2\}}\:d^3\vec{r}}$, with $\epsilon$ as the position dependent dielectric constant. For this cavity, the volume has a value of $V_{mode}=0.96\frac{\lambda^{3}_{c}}{n^3}$, where n is the index in AlGaAs. 
The term $F_{PC}$ describes the modification of the SE rate due to the presence of the photonic crystal lattice and modes other than the cavity mode. This modification results in a suppression of emission \cite{ref:Lee2000Fpc,ref:Dirk2005}. The cavity mode volume can be quite accurately derived from FDTD calculations, and with the cavity Q derived from the spectra, we calculate an  $F_{C} \approx 30$. We can also derive an average value of the suppression $F_{PC}$ from the experimental data. The total intensity collected by the spectrometer $I_{\phi}$ is given by the emission rate \ref{eq:Fcav}, integrated over the spatial and spectral density of the emitters.
\begin{equation}
\label{eq:Iphi} I_{\phi} = \Gamma_{0}\int{ d\vec{\mu}}\int{d\lambda }\int{d^3\vec{r}\left\{\eta_{C} F_{C} \cos(\phi)\left(\frac{\vec{E}\cdot\vec{\mu}}{\left|\max\{\vec{E}\}\right|\left|\vec{\mu}\right|}\right)^2 \frac{\Delta\lambda_c^2}{\Delta\lambda_c^2+4(\lambda-\lambda_c)^2}+\eta_{0} F_{PC}\right\}\rho{(\lambda,\vec{r},\vec{\mu})}}
\end{equation}

Here $\phi$ is the value of the polarizer angle, $\eta_{C}$ is the collection efficiency due to the cavity radiation profile, $\eta_{0}$ is the collection efficiency from an emitter embedded in the PMMA layer and uncoupled to the cavity, $\rho{(\lambda,\vec{r},\vec{\mu})}$ contains the spectral and spatial distribution of the dots and the orientations of their dipole moments. The value of $\eta_{C}$ can be estimated from numerically integrating the PC emission profile over the numerical aperture (NA) of the lens, and is $\eta_{C} \approx 8\%$. The coupling efficiency of the bulk emitter is given by the integral over the sub-critical solid angle defined by the NA of the objective lens $\eta_{0} = \frac{1}{4\pi} \int^{2\pi}_{0} {d\phi} \int^{arcsin(\frac{NA}{n_{PMMA}})}_{0}{sin(\theta)d\theta} \approx 3\%$ (Here $NA=0.5$ adn $n_{PMMA}=1.5$). Since the cavity modes are primarily linearly polarized, the ratio of the integrated intensities for polarizer angles of $0^o$ and $90^o$ gives:  
\begin{equation}
\label{eq:R} R=\frac{I_{0^o}}{I_{90^o}} \approx F_{C}
\frac{\eta_{C}}{\eta_{0}}
\frac{
\int{ d\vec{\mu}\: d\lambda\: d^3\vec{r}\: \rho{(\lambda,\vec{r},\vec{\mu})} \left(\frac{\vec{E}\cdot\vec{\mu}}{\left|\max\{\vec{E}\}\right|\left|\vec{\mu}\right|}\right)^2 \frac{\Delta\lambda_c^2}{\Delta\lambda_c^2+4(\lambda-\lambda_c)^2}}
}{
\int{ d\vec{\mu}\: d\lambda \: d^3\vec{r}\: F_{PC}\:\rho{(\lambda,\vec{r},\vec{\mu})}} 
}
+1
\end{equation}

Using the value of $F_C$ and the x-dipole field, the average value of $F_{PC}$ over the lattice around the cavity can thus be estimated from the spectra. Assuming random dipole orientation, the integral over $\vec{\mu}$ gives $\frac{1}{3}$. We take the spectral distribution of the dots to be Gaussian and centered at 850 and 950 nm with a FWHM of 100 nm. The cavity lineshape is Lorentzian with a FWHM of 2 nm. Only dots which are excited by the pump are contributing to the intensity, and so the spatial density corresponds to the excitation of a uniform dot distribution by a Gaussian pulse with a FWHM of 600nm in the x,y plane and uniform in z, since the layer is thin relative to the focal depth. The integral over the field is done numerically with the simulated cavity field components. The spectral data shown in Fig. \ref{fig:polspec} b. gives $R \approx 1.07 $. Using the value of $F_C = 30$, $F_{PC}$ is estimated to be  $F_{PC} \approx 0.6 $, which qualitatively agrees with the values found in \cite{ref:Lee2000Fpc,ref:Dirk2005}.  

In conclusion, we have shown coupling of PbS quantum dots dissolved in PMMA to photonic crystal cavities at room temperature. The dot emission maps out the cavity resonances and is enhanced relative to the bulk emission by a maximal Purcell factor of 30. Using the emission spectra with and without cavity lines, we derive a photonic crystal lattice spontaneous emission rate modification of $\approx 0.6$. To our knowledge this is the first demonstration of coupling of colloidal quantum dots to photonic crystal cavities, and the fist use of such dots as a broadband on-chip source for cavity characterization. The deposition technique presented in this work is easy to use and allows many different types of emitters to be deposited and re-deposited onto both passive and doped cavities. E-beam lithography can further be used to remove doped PMMA from regions around the cavity, and increase the probability of receiving the signal from a single emitter spatially aligned with the cavity field maximum.    
\acknowledgements{
Financial support was provided by the MURI Center for photonic quantum information systems (ARO/ARDA
Program DAAD19-03-1-0199), and the Stanford OTL Research Incentive. Ilya Fushman, is supported by a training grant from the National Institute of Health. Dirk Englund is supported by the NDSEG fellowship}

\clearpage

\begin{center}
\begin{bf}
List of Figures
\end{bf}
\end{center}

\begin{itemize}
\item
\begin{bf}
Fig. 1.\end{bf} PbS quantum dot spectra: 850 and 950 dot spectra taken on a bulk silicon wafer.   
\item
\begin{bf}
Fig. 2.\end{bf} Left: Scanning electron micrograph showing the photonic crystal cavity (a). Middle: Simulated electric field intensity of the x (b) and y (c) dipole modes in the asymmetric cavity. The measured Q factors are \~ 400 and 200 respectively.
\item
\begin{bf}
Fig. 3.\end{bf} Cavity resonances mapped out by quantum dots in PMMA. Left: Polarization dependence of modes confirming that they are x and y dipole modes. Right: Ex dipole mode measured at two orthogonal polarizations. Angles refer to analyzer positions.

\end{itemize}
\clearpage

\clearpage

\bibliographystyle{apsrev}
\bibliography{PBS1a}

\end{document}